# Intrinsic feature between malignant tumor cells and human normal leukocytes with statistical decision tree analysis via Raman spectroscopy

Yixin Dai,[1] Wenxue Li,[1] Liu Wang,[2] Chuan Luo,[3] Qing Huang,[2] and Lin Pang[1,*]

[1]College of Physics, Sichuan University, Chengdu, China
[2]Department of Laboratory Medicine, Daping Hospital, Chongqing, China
[3]Department of Laboratory Medicine, Southwest Hospital, Chongqing, China
*Corresponding: panglin_p@yahoo.com

Abstract: In this study, the combination of a developing data mining technique called statistical decision tree analysis method and Raman spectroscopy was proposed to differentiate human normal leukocytes from malignant tumor cells. Statistical results obtained indicate this method possesses an admirable performance of a mean classification accuracy of 94.43% on the one hand, base adenine and amide I are recognized as potential characterizations of main- and sub-intrinsic biological difference in between on the other hand. Moreover, these certain Raman bands reflecting intrinsic physiological differences can be directionally extracted from whole fingerprint spectra and then provide a fast and accurate manipulation for spectrum identification.

1.  Introduction

Cancer is one of the most threatening diseases to human life and health [1]. Tumors can originate from the pathology of almost every tissue or cell of body. Once the malignant tumors develop and they can overly invade the surrounding tissue and proliferate cells via body fluids such as lymph or blood [2,3]. Hence, the developments of fast detection and corresponding diagnosis technique for diverse cancer cells have been widely reported in recent years [4-6]. The use of spectroscopic techniques such as fluorescence, infrared and Raman are obtaining importance in clinic cancer research since they can provide fast molecular-level information with minimal invasive damage [7-9]. Raman spectroscopy exploits the frequency shift, which occurs when samples are illuminated with laser light owing to inelastic scattering of photons by bond vibrations of molecular constituents, has been proven extremely versatile and has guided extensive applications in many fields [10,11]. Especially Raman spectroscopy is an advanced way to probe biochemical changes in histological and cytological samples [12-15]. Label-free and non-destructive can specifically reveal characterizations in fingerprint regions and offer high possibility for detection down to molecule level. But the spectral similarities obtained from different species make it difficult to differentiate categories just based on these spectral similarities.

The principal component analysis- linear discriminant analysis (PCA-LDA) and partial least-squares-discriminant analysis (PLS-DA) have been employed to classify tumorous and normal cells [16,17]. Some data mining techniques, for example, decision tree, random forest and convolutional neural network (CNN) have been also reported for classification [18-20]. Although these techniques can achieve quite excellent classification effects, they just offer a convenient strategy for classifying different types of cells or molecules rather than determining intrinsic biological difference of which at biochemical molecule function group level. Therefore, other statistical analysis methods which are capable of searching essential difference carriers that cause significant spectrum difference, needs to be developed urgently.

In this paper, a data mining technique analysis method, named statistical decision tree (SDT), was proposed and used to evaluate binary classification capability of human normal leukocytes and malignant tumor cells based on their Raman cell spectra, as well as to attempt to find one or several peaks and corresponding biochemical molecule function groups reflecting intrinsic biological differences in between. The desirable purpose is to implement more precise Raman spectra identification of tumor and normal cells after these potential intrinsic differences were found, in addition to examine the feasibility of establishing an efficient and fast clinical diagnosis method based on SDT analysis and further cancer early screening. Results obtained demonstrate that not only SDT possesses a powerful performance with a mean classification accuracy of 94.43% for Raman cell spectra of human normal leukocytes and malignant tumor cells, but also the principal intrinsic biological feature differences may be characterized by specific changes of base adenine and amide I. It is believed that SDT analysis method will be possessing great potentials in wide applications, such as predicting the prognosis and overall survival rate of cancer patients even selecting effective therapy method.

## 2. Experiment section

### 2.1 Cell preparation

Leukocytes were isolated from peripheral blood (PB) of healthy donors with informed contents according to school of Medicine Ethics Committee. All experiments were performed in compliance with the relevant laws and institutional guidelines. Leukocyte samples were obtained from four heathy donors. Leukocytes were isolated by following immunomagnetic negative selection [21]. A total of 150 mL of fresh venous blood was drawn by venipuncture of elbow and then immediately collected into anticoagulant blood collection tubes respectively (heparin for lymphocytes, EDTA for residual leukocytes) to isolate white blood cells according to manufacturer's protocol. The acute leukemia cells were grown in culture with RPMI 1640 medium. The carcinomas in certain tissue parts, all of which were derived from tumor cells and grown in Dulbecco's Modified Eagle's Medium (DMEM) plus 10% FBS. The adherent cells were harvested with trypsin and resuspended in PBS buffer. All cell lines were cultivated at 20 °C in a humidified atmosphere containing 5% $CO_2$. All cells were washed and resuspended with PBS buffer for the further Raman spectroscopic analysis.

### 2.2 Raman spectroscopy measurement

Raman spectra were acquired in the spectral range of 600-1700 $cm^{-1}$ using a confocal Raman spectrometer platform of Raman-AFM system (Horiba JY HR evolution, France) equipped with an Olympus optical microscope and a charge-coupled device (CCD) detector. A 785 nm near-infrared laser was focused on samples with a 50x objective (N.A. =0.5) in a confocal arrangement. The laser power on the sample was about 50 mW. Lab-Spec 6 software was used for Raman data acquisition and analysis and the spatial resolution was less than 0.5$cm^{-1}$. All Raman measurements were completed in batches under the same conditions. A total of 749 leukocytes spectra and 365 malignant tumor cells spectra (145 renal cancer cells, 60 breast cancer cells and 160 leukemia cells) were obtained respectively after measurement. Three types of cancerous cells were not further categorized and labeled in this paper, thereby all measured spectra of cancerous cells were regarded as a single collection of malignant tumor cells.

### 2.3 Data analysis and statistical decision tree analysis method

A polynomial baseline correction was applied to subtract spontaneous fluorescence background noise for all measured raw spectra. Then all background-subtracted Raman spectra were smoothed and normalized. To reduce volume of follow-up input data into model, improve operation speed and avoid misalignment caused by possible peak shift, 17 characteristic Raman peak positions (623, 646, 728, 748, 783, 853, 936, 1004, 1033, 1099, 1126, 1211, 1256, 1303, 1446, 1658 $cm^{-1}$) for each spectrum based on the average spectrum of all measured cell spectra

were fixed in advance and then corresponding spectral intensities were extracted and zero-mean (z-score) normalized for each processed spectrum. Therefore, the 1113 rows by 18 columns matrix of 17 columns feature attributes (characteristic peak intensity) with a column of label information of 1113 spectrum were converted as the input data of SDT analysis.

The traditional decision tree is a kind of multivariate supervised machine learning algorithm with becoming increasingly popular. The basic model structure based on classical ID3 (Iterative Dichotomiser 3) is iterative [22]. The tree is constructed by recursively separating the total sample set in branches and more specifically, splitting feature attributes of sample set into two adjacent subgroups which is repeated at each internal node until the very bottom layer leaf nodes conditions are satisfied. Information gain was used to control tree formation and ensure achievement of the maximum information gain between root and leaf node for the purpose of constructing an optimal classifier. In each internal node, one of feature attributes of sample set is selected and then sample set is divided [23]. Here every feature attribute denotes a Raman band of certain biochemical molecule function groups carrying characteristic vibration information. In order to avoid the over-fitting, max-depth of tree and min-impurity-decrease and other parameters are automatically optimized in pruning. The traditional decision tree is an appropriate discrimination method but lacks of sensitivity for great similarity, especially for structurally similar molecules containing diverse function groups or homogeneity of different types of cells. To avoid such defect, our designed SDT can generate plenty of ID3 base tree classifiers successively. In this study, the number of trees is set to 500. For a generated tree, database is blindly and randomly divided into training set and validation set at a 4:1 ratio. In this paper, each tree was generated from different training set, therein the advantages of using a different training set for each tree generated are to sufficiently retrieve characteristics of total sample set as far as possible and to avoid model becoming biased with increase of the number of tree classifiers. Internal root node and secondary nodes as primary controllers of subsequent tree formation directly represent the main- and sub-intrinsic feature attributes of sample set respectively, wherein the statistics of all internal node distributions can be utilized to evaluate the feature difference level of a biochemical function group in classification of human normal leukocytes and malignant tumor cells. Hence, the issue of how significant are these biological differences across various types of cells can be solved by statistical node distributions.

Computer operating system is Windows 10, the central processing unit (CPU), the Read-Only Memory (ROM) and the Random-Access Memory (RAM) are Intel Core i3-3770 (3.40GHz), 1TB and 8GB respectively. All statistical decision tree analysis calculations were accomplished in Spyder based on Annaconda 3.

## 3. Results and analyses
*3.1 Raman spectra of malignant tumor cells and leukocytes*

To visualize the spectral differences of human normal leukocytes and malignant tumor cells, the Raman spectra in the fingerprint range of 600-1700 $cm^{-1}$ are described as shown in Fig. 1, in which 17 selected and fixed Raman peak positions reflecting biochemical molecular contributions from various cellular constituents are also marked. Their assignments have been summarized by extensive researches [24-27].

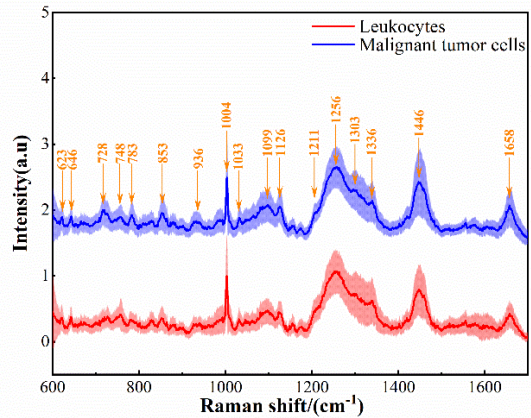

Fig. 1 The Raman spectra of leukocytes and malignant tumor cells. The color solid lines and the shaded lines denote average Raman spectra and standard deviations, respectively. 17 preset and fixed peak positions are marked by orange arrows.

Actually the spectra of tumor cells and leukocytes appear very similar since they share many bands that can be assigned to cellular constituents, including nucleic acid, proteins and lipids [24]. Spectra in the region of 700-800 cm$^{-1}$ are mainly dominated by characters of nucleic acid, like ring breathing vibration of nucleic acid bases at 748 (thymine),783 (thymine, cytosine, uracil) and adenine at 728 cm$^{-1}$. The average spectral intensities of Raman peak at 728 cm$^{-1}$ in tumor cells are a little stronger than that in leukocytes, suggesting more nucleotides are possibly found in tumors due to their larger nuclei containing higher nucleic acid concentration and loosely chromatin, as well as the cancerization could induce nucleic acid changes [28-30]. On account of large laser probe spot, the cytoplasm surrounding the nucleus and compositions of membrane were also partly detected. Small aromatic amino acid bands of phenylalanine and tyrosine show up around 623 and 646 cm$^{-1}$. The broad feature around 1336 cm$^{-1}$ arises from C-H deformation of most of cellular compositions and C-N stretching vibration shows up around 1099 cm$^{-1}$. Further vibrational bands of aromatic amino acids are also found at 1004 and 1033 cm$^{-1}$ (phenylalanine), 853 cm$^{-1}$ (tyrosine). The broad band around 1256 cm$^{-1}$ is originated from different cellular protein secondary structure [12]. Also, the coincidence of amide III band around 1211 and 1256 cm$^{-1}$ and amide I band around 1658 cm$^{-1}$ expresses different protein secondary structure conformation alterations. Similarly, in protein distinctive bands, the average spectral intensities of tumors are more intense than that in leukocytes. Average cell spectra are a little different from each other, however, the intensity difference of each particular peak is still disparate which produces many difficulties for precise classification and extraction of main-intrinsic biological feature difference. Therefore, it is necessary to introduce data mining technique such as SDT analysis method into Raman cell spectra to resolve this problem.

*3.2 Results of statistical decision tree analysis method*

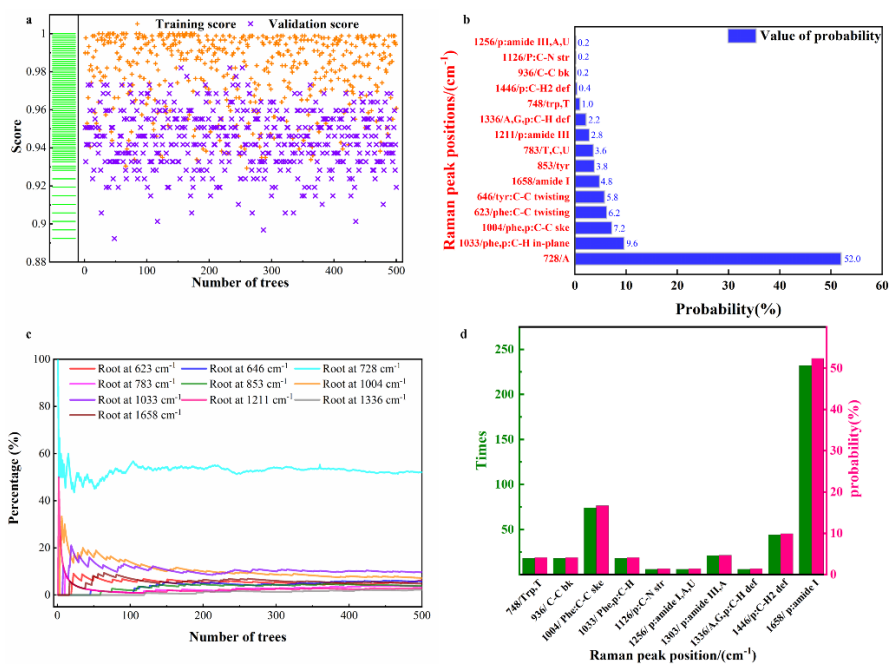

Fig. 2(a) The overall score distribution of training and validation set in 500 calculations. Training score by orange positive sign and validation score by purple cross sign represent the cognitive ability for these two types of cells Raman spectra and an overall prediction accuracy for each unlabeled spectrum respectively. (b) Statistical probability distribution of 15 root nodes in generated trees, every root node denotes a Raman peak and corresponding biochemical molecule function group. The probability was computed by ratio of number of one Raman peak serving as root to total trees (n=500). Two of 17 selected Raman peak positions were not appearing as root (*i.e.* their probabilities were zero). (c) Percentage change trends of ten Raman peaks with higher probability along with increase of base trees. (d) When root of tree was at 728 cm$^{-1}$, the number of times that ten Raman peaks emerged as internal secondary node and their corresponding probabilities were shown at length. The green denotes emerging times of individual Raman peak serving as internal secondary node and the pink denotes corresponding probabilities. Abbreviations: Phe, phenylalanine; Tyr, tyrosine; Trp, tryptophan; A, adenine; T, thymine; U, uracil; C, cytosine; p, protein; bk, backbone; ske, skeletal; def, deformation; str, stretching;

In line with above-mentioned analysis method introduction, 500 ID3 tree classifiers were successfully generated and each base tree classifier was implemented under the same conditions. Fig. 2(a) shows the detailed scatter distribution of each training score and validation score in sequential 500 times calculations, in which training score varied from 0.92 to 1.0 and validation score varied from 0.892 to 0.982 correspondingly, indicating the precise binary classification of normal and cancer cells could be realized. Every calculation can generate an independent tree classifier, where root contains all the class labels and reflects purity of sample integration so that it better controls subsequent formation. In tree, root node and secondary node as primary controller of regulating following growth deliver the main- and sub-intrinsic feature difference of sample set. The statistics of root in all trees were used to evaluate the feature difference level for each biochemical molecule function group as shown in Fig. 2(b). Raman peak position at 728 cm$^{-1}$ assigned to base adenine achieved the highest probability of more than 50% while probabilities of other Raman peak positions were lower than 10%, which illustrates adenine perhaps possesses the greatest influence on classification of malignancies and leukocytes. Meanwhile, percentage change trends of ten Raman peaks with higher probabilities along with increase of calculation times were plotted in Fig. 2(c). Adenine still kept the highest proportion, it proves the correctness of statistical results. Adenine is thus determined as the main-intrinsic biological feature difference among human normal leukocytes and malignant tumor cells. Some medical studies have revealed DNA conformation change may be connected with the band intensity differences of base adenine around 728 cm$^{-1}$ between tumor and normal cells, as well as alterations in the level of nucleic acid are associated with tumor burden and malignant progression [31,32]. Besides rapidly proliferating tumor cells need extensive ATP generations to maintain energy status and increase biosynthesis of macromolecules due to core cellular metabolism and basic needs for dividing cells [33]. Appearances of aromatic amino acid peaks reflect components of proteins in cells, such as peaks at 1004 and 1033 cm$^{-1}$ whose probabilities achieved 9.6% and 5.8% respectively, suggesting the intrinsic differentiation characterizations of normal and cancerous cells are affected by protein constituents as well. Higher expression of aromatic amino acids bands serving as root node may be related to a fact of benign and malignant degree of tumor [34]. And the coincidence of amide I and III reflects divergent protein secondary structure conformation and their bands are extremely sensitive to subtle changes in the protein secondary structure [35]. Specifically, the spectral intensity differences of amide bands among tumor and normal cells are closely related to a fact that α-helix of stable protein conformation would transform into turbulent β-sheet conformation [25,31,36].

However, a series of physiological activities of tumor cells including carcinoma origin, proliferation and metastasis, as well as reasons of causing the histopathological difference expressions of tumor cells and normal tissue cells are excessively complicated. The intrinsic biological feature differences of human normal leukocytes and malignant tumor cells cannot be completely characterized by adenine specific alteration. Therefore, relevant statistics of all secondary nodes of 500 trees were also investigated when root node was attributed to adenine molecule function group of Raman peak at 728 cm$^{-1}$ as shown in Fig. 2(d). Therein amide I Raman peak at 1658 cm$^{-1}$ serving as secondary node in all generated trees, emerged 232 times and its probability was 52.4% (232 out of 443), that demonstrates specific alteration of cellular amide I reflects sub-intrinsic biological difference between tumor and normal cells. It also means the intrinsic biological discrepancies of human normal leukocytes and malignant tumor cells are induced by variations of adenine and amide I conjointly.

## 4. Discussion

The spatial resolution used by different Raman system would really affect the data repeatability or reproducibility. If the spatial resolution of the system is much higher than the size of cell, the Raman signal would depend on the cellular position as reported in the reference [37], in which a minimum sample size of Raman spectra can be used to replace the full hyperspectral Raman image to achieve discrimination between different cell populations. There was also reported in which several regions within a single live cell were probed at random locations to evaluate the variation, however, there were no noticed spatial variations [38]. To evaluate our Raman-AFM system setup, the spectral scanning was conducted across different locations of a monocyte cell in a manner from edge-center-edge, the remarkable spectral differences were not detected [39]. This means that our measurement system does not possess enough spatial resolution to distinguish the cell locations.

On the other hand, there is an actual data repeatability issue regarding to the measurements, which results from the variations of the illumination power, integration time and collection efficiency of the measurement system, leading to the Raman intensity irreproducibility or inconsistence or differences in terms of the absolute values. However, the Raman signature of the matter would not depend on the absolute spectral values, but the related relationship or relative intensities among all the spectral values. The evaluations show that the normalized spectral values nearly keep the same although there were small variations at different acquired cell positions.

The proposed SDT analysis is a multivariate statistical analysis method to extract the underlying intrinsic biological difference, while the traditional decision tree model (1X) and other discrimination methods such as principle component analysis- linear discriminant analysis (PCA-LDA) and support vector machine (SVM) have been conducted to give reasonable classification performance. As a statistic model, the data sets, in principle, would be extremely large to provide accurate predictions. Specifically, each base tree classifier was utilized to implement precise Raman cell spectra classification and the statistical probability distributions of different Raman bands as internal node were used to extract the intrinsic biological difference. The number of SDT trees is not critical as long as the performance converges. As shown in Fig. 2(c), 200 trees classifiers would be enough to just ensure the validity of statistics. The number of trees could be properly reduced in the many applications, thereby improving the computing speed. These preliminary results show the potential use of proposed approach to establish a better understanding of the underlying biochemical molecule difference between diverse cell populations.

In this paper, the intensities of 17 characteristic Raman peak positions were selected as feature attribute to input subsequent SDT analysis. Although less peaks as input feature attributes for sample sets could potentially improve computing speed, the processing becomes unstable, leading to classification accuracy significantly drop. To improve the proposed SDT analysis, the statistic nature of the model would be further explored while combining all the variable variations in the future. Furthermore, SDT analysis will be conducted in different bio/medical fields with extremely weak but intrinsic differences, such as featuring certain malignant tumor cells from normal tissue counterparts, various malignant tumor cells and different phenotypes of same types of tumor cells.

The main- and sub-intrinsic biological feature difference between human normal leukocytes and malignant tumor cells has been extracted from all measured individual cell spectra. In order to further examine SDT analysis method's validity, local mean Raman spectra and corresponding standard deviation of two kinds of all measured cells at region of 700-780 $cm^{-1}$ and 1640-1700 $cm^{-1}$ are depicted as shown in Fig. 3.

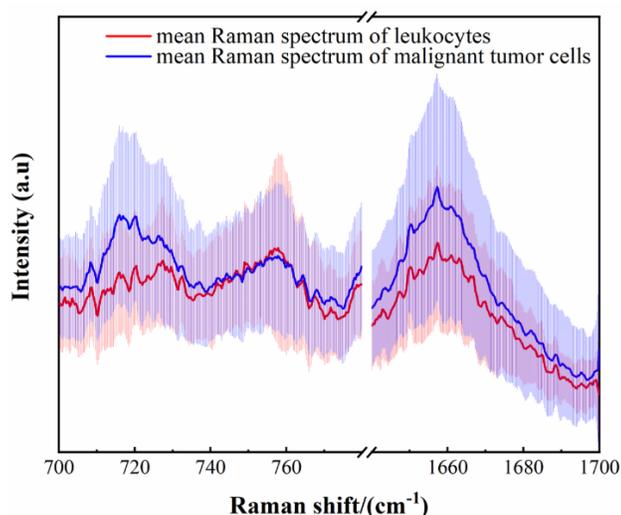

Fig. 3 The local Raman spectra of malignant tumor cells and leukocytes in the regions of 700-780 cm$^{-1}$ and 1640-1700 cm$^{-1}$. Color solid lines denote average Raman spectra of two types of cells. The shadow denotes standard deviation.

In consideration of mean cell spectra, spectral intensities of peaks at 728 and 1658 cm$^{-1}$ between cancerous and normal cells are really existing a few weak differences. However, the measured spectral intensity is easily affected by measurement environment changes such as cell activity or morphology even different measurement parameters, which can inevitably lead to large deviations for spectral intensity. Although mean spectra can provide some characteristic information across diverse cell species, it also obliterates individual difference at the same time. Thus, using mean Raman spectra to conduct practical qualitative or quantitative analysis is not unreliable. Two molecule function groups of base adenine and amide I reflecting intrinsic biological feature difference between leukocytes and malignancies can be extracted by internal node probability distribution, besides other spectral peaks with minor differentiations would be taken into consideration for SDT analysis method. SDT results obtained show exact consistency with cell species spectra as well. Furthermore, SDT analysis does not depend on absolute intensity comparison of certain bands any more, but provides a statistic strategy to interpret the underlying sources of classification to extract physiological meaning. On account of these characteristics, it is believed that people can directionally scan certain peaks instead of inherent signal intensity level comparison to perform more reliable qualitative and quantitative analysis and further to offer assistance to fast and real-time detection, especially beneficial for CTCs technique. Maybe the most important contribution is to provide constructive suggestions about issues of tumorigenesis and metastasis of carcinomas by those molecule function group carriers with intrinsic biological feature differences.

## 5. Conclusions

This study demonstrates the combination of proposed SDT analysis method and traditional Raman cell spectra is appropriate afterward it has been successfully applied in classification of human normal leukocytes and malignant tumor cells. Results obtained suggest this integrated technique not only achieved a mean classification accuracy of 94.43%, but also base adenine and amide I were recognized as main- and sub-intrinsic biomedical feature differences among cancer and normal. The advantages of abandoning inherent signal alterations of some certain bands and assessing which Raman band carries the most abundant difference information with

SDT analysis method have been proven to be desirable since it is competent to extract important feature attributes from crowed sample set. Even more importantly the integrated data mining technique will possess potential applications in clinic, such as identification of cell biomarker in diagnostic pathology that are not visible to the naked eye and fast and accurate cancer early screening by detecting blood or other body fluids component changes coupled with CTCs technique.

**Funding**

National Natural Science Foundation of China (NSFC, 61377054, 61675140); The Scientific Foundation of Southwest Hospital (SWH2016LHYS-01);

**Disclosures**

The authors declare that there are no conflicts of interest related to this article.